\documentstyle[11pt,epsfig,epsf]{article} 
\newcommand{\bea}{\begin{eqnarray}}
\newcommand{\eea}{\end{eqnarray}}

\begin{document}
%THE TEXT STARTS HERE
\begin{flushright}
{\large HIP-2003-35/TH}\\ 
{hep-ph/yymmddd}\\
\end{flushright}

\begin{center}

{\Large\bf Hunting Radions at Linear Colliders}\\[20mm]

Anindya Datta \footnote{E-mail: datta@pcu.helsinki.fi},
Katri Huitu \footnote{E-mail: huitu@pcu.helsinki.fi} \\
{\em Division of High Energy Physics, Department of Physical Sciences, and\\
Helsinki Institute of Physics,\\
P.O. Box 64, 
%Gustaf H\"{a}llstr\"{o}min Katu 2, Helsinki 00014, 
FIN-00014 University of Helsinki, Finland}
\end{center}

\vskip 20pt
\begin{abstract}

We investigate the possibility to disentangle the radions in the
Randall-Sundrum scenario from higgs boson at the next generation high
energy $e^+ e^-$ linear colliders.  Due to trace anomaly, the radion
coupling (and in turn the branching ratio) to gluons is enhanced over
the same for the Standard Model (SM) higgs. We study the radion $\phi$
production at electron-positron colliders, via the process 
$e^+e^-\rightarrow \nu\bar\nu \phi$,
and propose to investigate radion decay
to a pair of gluons. At 500 GeV center of mass energy, our signal is
not very promising. We find our results are encouraging for finding
radion and differentiating it from higgs for center of mass energies
of 1 TeV and above.

\end{abstract}

\vskip 1 true cm

%\begin{flushleft} {\bf Introduction} 
%\end{flushleft}
\noindent
Recently proposed models \cite{add,rs}, aiming to cure the hierarchy
problem between electroweak (EW) and Planck scale, offer an exciting
possibility to test the gravitational interactions at the TeV
colliders. The models in \cite{add,rs} require our universe to 
have $(3 + n)
+ 1$ space-time dimensions, of which the extra, $n$, space-like dimensions
are compactified. The ADD \cite{add} model requires
relatively large compactification radius ($\sim 1~mm$). In the
Randall-Sundrum (RS) scenario \cite{rs}, which is the focus of this
analysis, there is only one extra space-like dimension compactified on
an $S_1/Z_2$ orbifold. Unlike the ADD, RS-scenario does not require a
large compactification radius for the extra compactified space-like
dimension. The radius of compactification is of the order of the
Planck length, and interestingly it is a dynamical object. Moreover, the
assumed geometry of space-time is non-factorisable in the sense that
$3 + 1$ dimensional metric is scaled by an exponential warp-factor
depending on the extra space-like dimension. Apart from the Kaluza-Klein
tower of gravitons, the low
energy effective theory has also a graviscalar, called radion.
Goldberger and Wise \cite{gold} have proposed a method to generate a
potential for this scalar, by introducing a scalar field in full five
dimensional manifold. This in turn dynamically generates a
vacuum expectation value (VEV), $\Lambda_\phi$, for the radion.
Following the proposal in \cite{gold}, $\Lambda_\phi$ comes out to be
of the order of TeV without finetuning the parameters. 
The Kaluza-Klein excitations of the bulk fields 
are of the order of a few times TeV \cite{gold}.
Assuming a stabilization similar to \cite{gold}, the radion is 
likely to be the lowest lying gravitational state.
Here we will assume that only radion is of interest in the studied
experimental situations.

Radion $\phi$ couples to the Standard Model particles in a model independent
fashion via the trace of the energy momentum tensor, 
\begin{equation}
L_{int} = \frac{1}{\Lambda_\phi}\;T^\mu _\mu\; \phi.
\label{radion_int}
\end{equation}
This implies that radion interaction with SM fields is very similar
to the higgs boson, but with a suppressed strength depending on the
value of $\Lambda_\phi$. Apart from
eq. (\ref{radion_int}), radion can have  a mixing with the SM higgs
via the following term in the action:
\begin{equation}
S=-\xi \int d^4 x \sqrt{-g_{vis}} R(g_{vis}) H^\dagger H.
\label{radion_mix}
\end{equation} 
Here the Ricci scalar $R\;(g_{vis})$ corresponds to the induced four
dimensional metric, $g_{vis}$, on the visible brane, $H$ is the
electroweak higgs boson, and $\xi $ is the mixing parameter.

Phenomenology of radion has been studied extensively in recent
literature in the context of collider experiments like LHC, a
linear collider, or
in the cases like muon $(g-2)$, $K^0 - \bar K^0$ mixing, and
electroweak $\rho$-parameter \cite{radion_pheno}. 
Effects of radion have also been
studied in the context of unitarity in gauge boson scattering
\cite{unitary}. While the low energy experiments try to constrain the radion
parameters, collider studies are naturally more focussed on the
possible search strategies of this particle at the planned
experiments.  From the collider studies it is evident that LHC would
be the best place to discover/exclude this kind of a scalar if the
radion vev is in the ball park of a TeV or so.

At the same time, in order to make correct theoretical deductions, it
is crucial to know the real identity of a scalar detected in
an experiment. 
In this work we consider the detection and identification of radion,
which may be difficult, since the decay modes of radion and higgs
are identical.
A recent study \cite{cheung} aimed at the high energy $e^+ e^-$
collider, proposes associated production of a radion with a SM higgs
mediated by KK gravitons. This particular final state is an outcome of
higgs-radion mixing. Higgs-radion mixing have also been investigated 
to study the complementarity of higgs/radion signals at 
$e^+ e^-$ colliders \cite{gunion}.
In the
following we will study the separation of the radion signal from the
higgs signal at a linear collider, with varying radion-higgs mixing
parameter.

As mentioned earlier, radion couples to the SM particles via the trace
of energy momentum tensor.  Being massless, gluons and photons cannot
couple to a radion at classical level. At quantum level they do 
couple to radion, with a coupling proportional to the $\beta$-function
coefficients of $U(1)_Y$ and $SU(3)_C$, respectively. It turns out that
for $\Lambda_\phi \sim 1$ TeV, radion couplings to a pair of gluons or
photons are enhanced over the corresponding couplings of higgs. This
was the key feature of all the previous exercises looking for a  radion at
hadronic machines. At an $e^+e^-$ collider, we propose to produce
radions in the process
$e^+e^- \rightarrow \nu \bar \nu \phi$, and to
study radion decay to gluon gluon. Though radion branching ratio
to gluon gluon is quite sizeable, this decay channel cannot be used 
at hadron colliders due to overwhelmingly large QCD background.
At  an $e^+e^-$ machine, one can hope to cope with two jets with missing
energy-momentum final state.

After a brief discussion of higgs-radion mixing, we will continue with the 
signal and background analysis.   
The action in eq. (\ref{radion_mix}) leads to a term bi-linear 
in curvature and higgs in the Lagrangian \cite{mixing,mixing_1},
$$
{\cal L} = - 6\xi \Omega^2\left(\Box \ln \Omega 
+ (\nabla \ln  \Omega)^2\right)  H^{\dagger} H ,
$$
\noindent where 
$$\Omega=e^{-(\gamma/v) \phi(x)},\; 
\gamma=v/\Lambda_\phi.$$ 
Clearly, this
induces a kinetic mixing between higgs and
the radion.  Additional mixing is introduced by the fact that both 
the neutral component of the higgs ($h$) as well as the radion ($\phi$) 
acquire VEVs ($\langle h\rangle \equiv v$).
To obtain fields with canonical quantization rules, it is necessary 
to make field redefinitions:
\begin{equation}
\pmatrix{ \phi \cr h \cr} \to \pmatrix{ \phi' \cr h' \cr} 
	\equiv Z_R \: {\cal M}^{-1} \pmatrix{ \phi \cr h \cr}  \ ,
\qquad
{\cal M} = \pmatrix{ \cos \theta  & -\sin \theta \cr
                     Z_R \sin \theta-6\xi \gamma \cos \theta 
                                 & Z_R \cos \theta +6 \xi \gamma \sin \theta
		    \cr}
\label{mixmatrix}
\end{equation}
where 
\begin{equation} \displaystyle
Z_R^2=1-6\xi \gamma^2 (1+6\xi) \quad {\rm and} \quad 
\tan 2 \theta = \frac{12 \xi \gamma Z_R  m_h^2}{m_h^2(Z_R^2-36\xi^2\gamma^2)-
m_{\phi}^2},
\label{mixing}
\end{equation}
where $m_h$ and $m_\phi$ are the higgs and radion mass parameters in
the Lagrangian. In the limit $\xi \rightarrow 0$, we recover back the
SM higgs from $h^{'}$.

 Requiring that the kinetic terms for the physical fields
$h'$ and  $\phi'$ be positive,  restricts us to \cite{mixing_1}:
\begin{equation}
\frac{-1}{12} \left( 1+\sqrt{1+4/\gamma^2} \right) \; \le \xi \; \le
\frac{1}{12} \left(\sqrt{1+4/\gamma^2}-1 \right).
\label{limits}
\end{equation}
For $\Lambda_\phi=1$ TeV, this translates to $-0.75 <\xi <0.59$. 
As seen from eq. (\ref{mixmatrix}), the mixing matrix of radion 
and higgs is not
unitary.  Therefore it is not always straightforward, which particle
should be called higgs and which should be called radion.  We
will always call $\phi^{'}$ radion and $h^{'}$ higgs in the following
calculations.

The above redefinition of the fields leads to the following
interactions between scalars and fermions or gauge bosons: 
\begin{equation}
{\cal L} = -\frac{1}{\Lambda_{\phi}} 
\left(
\sum_f m_{f} \bar{\psi}_f \psi_f - M_V^2 V_{A\mu}
V_{A}^{\mu}\right) \left[ a_{34} \frac{\Lambda_{\phi}}{v} h^{'} 
+ a_{12} \phi^{'}\right],
\label{rad_int}
\end{equation}
where $a_{34}$ and $a_{12}$ are defined in terms of the elements of the 
matrix $\cal M$:
$
a_{12} = \frac{1}{Z_R} \; \left(  {\cal M}_{11} + {\cal
M}_{21}/\gamma \right)$ and 
$a_{34} = \frac{1}{Z_R} \; \left( {\cal M}_{22} + \gamma {\cal M}_{12} \right)
$.

%----------------------------------------------------------------------
\begin{figure}[h]
\vspace*{.5cm}
\centerline{
\epsfxsize=8.cm\epsfysize=7cm\epsfbox{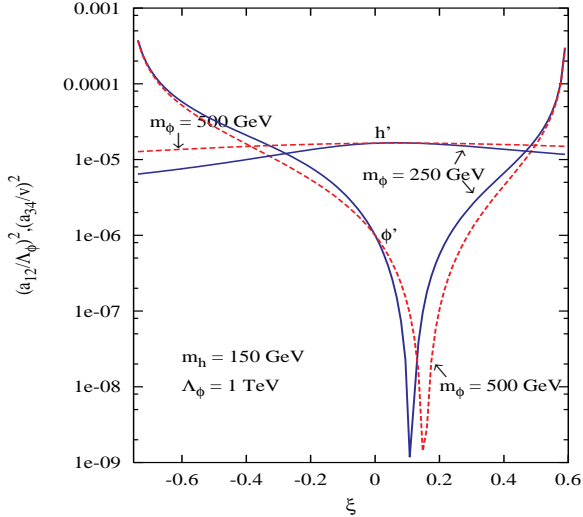}
}
\caption{Comparison of radion (marked $\phi'$) and higgs 
(marked $h'$) couplings
to the fermions or gauge bosons as defined in eq. (\ref{rad_int}).
The used mass parameters are $m_h=150$ GeV and $m_\phi=250,500$ GeV.}
\label{couplings}
\end{figure}
%--------------------------------------------------------------------

The effects of mixing on the higgs and radion couplings to fermions 
and gauge bosons are presented in fig. \ref{couplings}, see 
eq. (\ref{rad_int}) for the definitions of $(a_{34}/v)$ and
$(a_{12}/\Lambda_\phi)$. 
From the figure it is
evident that apart from the region around the special value of $\xi =
\frac{1}{6}$, radion couplings to gauge bosons or fermions are bigger
than the respective higgs couplings. 
Far enough from $\vert \xi \vert = \frac{1}{6}$, $\phi '$
couplings are enhanced over the $h '$ couplings. This phenomenon is
purely due to the nature of the higgs-radion mixing. While calculating
the radion (higgs) couplings, we have fixed $m_h=150$ GeV and
$m_\phi=250$ or 500 GeV.

The couplings to gluons can be written as follows \cite{laser}:
\begin{equation}
{\cal L}_{gg} =
\left[\frac{1}{\Lambda_{\phi}} \left(\frac{{\cal M}_{11}}{Z_R} b_3 - 
\frac{1}{2} a_{12} F_{1/2}(\tau_t)\right)
\phi^{'}+\frac{1}{v} \left(\gamma \frac{{\cal M}_{12}}{Z_R}  b_3 - 
\frac 12 a_{34} F_{1/2}
(\tau_t)\right) h^{'}\right] \frac{\alpha_{s}}{8\pi} G_{\mu\nu}^a G^{\mu\nu a}
\label{rgg}
\end{equation}
where $b_3$ is the QCD $\beta$-function coefficient.  $F_{1/2}$ is the
form factor from (heavy quark) loop effects. In each of these couplings the first 
term proportional to
$b_3$ is  coming from the trace anomaly.  We can see from eq. (\ref{rgg})
that the vertices higgs/radion -gluon-gluon have new
contributions, which change the production and decay of the higgs
boson. Couplings of radion/higgs to a pair of photons can also have 
anomalous contributions which we do not write here explicitly and
which can 
be found elsewhere \cite{laser}. We do not write either the trilinear 
couplings involving radion and a pair of higgs or vice-versa, as we are
not interested in these couplings in this work.   

\begin{figure}[h]
\vspace*{.5cm}
\centerline{
\epsfxsize=8.cm\epsfysize=7cm\epsfbox{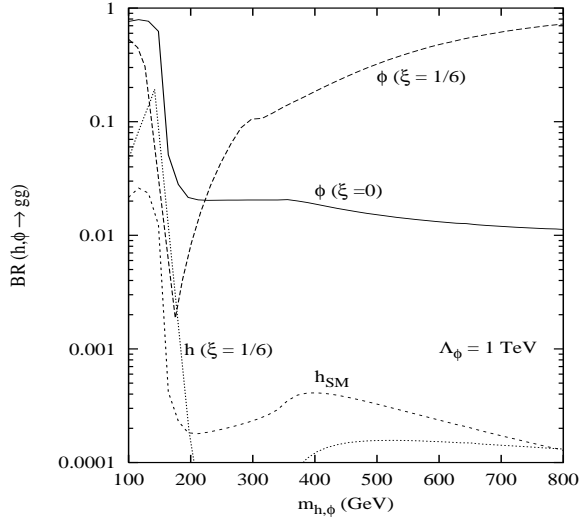}
}
\caption{Radion and higgs branching ratios for $\xi = 1/6$ and $\xi = 0$
for $\Lambda_\phi = 1$ ~TeV. Whenever required we fix the radion and 
higgs mass parameters at 150 GeV.}
\label{brgg}
\end{figure}

Without going into further details of the gluon couplings, we choose to
plot in fig. \ref{brgg} the radion branching ratio to gluons for both 
vanishing and
non-zero values of $\xi$. For purpose of comparison we also present
the same branching ratios for the higgs boson. For $\xi = 0$, $\phi
\rightarrow gg$ branching ratio is almost two orders of magnitude higher
than the same for $h$. The sharp fall of the branching ratio around
160 GeV can be accounted for by the opening up of $WW$ decay modes. The
radion branching ratio to a pair of gluons remains almost unchanged
with mass, once the $WW$ threshold is crossed.
For non-zero value of the mixing parameter the branching ratio rises 
with the radion mass. This is 
in sharp contrast with the higgs case. Higgs
branching ratio to gluons in presence of mixing tends to decrease
from its SM value for heavy $h^{'}$. 
 
%----------------------------------------------------------------------
%-------------------------------------------------------------------- 

What should we look for at a linear collider?
The lesson we have learnt from previous discussion is that the radion 
coupling to
a pair of gluons is enhanced due to the trace anomaly. This has been
exploited in previous studies on radion production at hadronic
colliders. The aim of the present work is to see the feasibility of
next generation high energy $e^+ e^-$ collider to differentiate
between the Randall-Sundrum type models with radions and
SM. As radion to photon photon coupling is also enhanced slightly (due
to the modest running of QED $\beta$-function), possibilities of producing
radions in photon photon collision (with back-scattered photons, using
laser and high-energy electron or positron beams) have been considered
\cite{laser}.  This cross-section is not very
impressive. Moreover, SM higgs has same kind of production and decay
channel with almost competing strength. Thus, this production mode may
not be suitable to differentiate between higgs and radion. This is why
we turn to the process $e^+e^- \rightarrow \nu
\bar \nu \phi (\rightarrow gg)$.  
We consider both the $WW$ fusion, which is the dominant production
mechanism for the linear collider c.m. energies, and radion strahlung,
$e^+e^-\rightarrow Z^*\rightarrow Z\phi$.
Two jets in the final state are
easier to deal with at an $e^+e^-$ machine than in a hadron collider.  
At the same time, radion
branching ratio to gluons does not fall steadily with its mass apart
from a sudden decrease around $2W/2Z$ threshold. This is related to
the fact that radion gluon gluon coupling is independent of radion mass
while for  higgs the corresponding coupling decreases rather sharply 
with increased higgs
mass. This is crucial in our analysis in the sense that unlike the higgs,
gluon gluon branching ratio of radion can be substantial at high
radion mass.  

Moreover, for $\xi = 0$, radion production via $WW$ fusion is
suppressed due to the factor $(v/\Lambda_\phi)$ in the coupling,
with
respect to higgs production in the same channel. Thus we have to be
careful to choose such a decay channel for radion that this
suppression factor can be overcome. This leads us to choose the radion
decay to a pair of gluons. Even for $\xi \neq 0$, we can see from
fig. \ref{couplings} that for a wide (allowed) range of $\xi$ and radion
mass, the radion coupling to a pair of $W$'s is larger than the same for
higgs. This in turn implies the higher radion cross-section even at
$\Lambda_\phi = 1$ ~TeV. On the other hand, when $\phi'-W-W$ coupling
is small (around $\xi = 1/6$), radion branching ratio to
gluon gluon becomes big. This compensates the suppression in
production.  
 This already proves the efficacy of
this particular channel.

We already mentioned that in the limit $\xi \rightarrow 0$,
$\phi'$ can be identified with the radion. In this work we will
discuss the detection of $\phi'$ through $gg$ decay.
We will not present the
other decay branching ratios of $\phi'$ here. These can be obtained from our
earlier works \cite{radion_pheno}.

We have calculated the cross-section for $e^+e^- \rightarrow \nu
\bar \nu \phi '$ (all three $\nu$-flavours are added appropriately)
and multiply by $\phi'$ branching ratio to a pair of gluons.  SM higgs
boson has this decay mode as well and produce similar kind of a
signal. On the other hand, higgs decay rate to gluons is
suppressed w.r.t radions as
discussed earlier. For the purpose of comparison we have also
calculated the two-jet + missing energy yield in $e^+e^-$ collision
via SM higgs production and decay. For low mass (around 100 GeV),
the SM higgs decays dominantly to a pair of b-quarks, thus also producing two
jets. If we assume a good b-jet identification/discrimination at
a linear collider and veto any b-jet, this gives us a handle to 
discriminate the SM
higgs from radion (around 100 GeV higgs/radion mass). We have also
estimated the cross-section of two jets + missing momentum final state
coming from $e^+e^- \rightarrow \nu \bar \nu \;(\gamma^\ast, Z^\ast, Z
\rightarrow) \;q \bar q $. The strategy is to compare the invariant mass
distribution of the jet pair. For radion one expects to have a peak
around the radion mass we are interested in. On the other hand, the
SM mass distribution of the jet pair has a continuum apart from a bump
around $Z$-mass due to the on-shell $Z$-production. Two-jet mass
distribution also peaks at low mass (of two jets), corresponding to
the soft singularity associated with the $q \bar q$-pairs fragmented
from a soft photon. We have used the following kinematic cuts on
signal and SM background.

\begin{itemize}
\item for two jets 
      $\vert p_T \vert > 15$ GeV and 
      $ |\eta_j|\le 3.0$. 

\item $\Delta r_{jj} \left( \equiv 
\sqrt{\Delta \eta _{jj} ^2 + \Delta \phi _{jj} ^2}\right)> 0.7$.

\item  ${p_T}\!\!\!\!\!\!\slash\; \;\;> 15$ GeV.
\end{itemize}

\begin{figure}[h]
\centerline{
\epsfxsize=8.cm\epsfysize=7cm\epsfbox{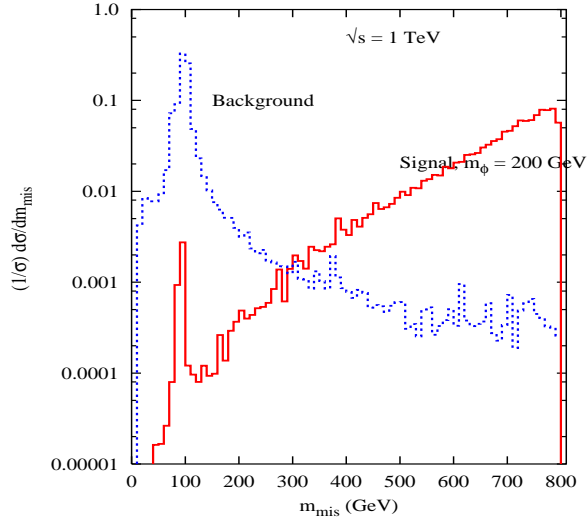}
}
\caption{Missing mass distribution of signal and background.}
\label{mis_mas}
\end{figure}

There is one more contrasting feature of signal and background which
helps us to reduce the background rate without affecting the signal
too much. There is a dominant part of the background in which the
neutrinos are coming from an on-shell $Z$. This can be easily seen
from the missing mass distribution of signal and background as plotted
in fig.
\ref{mis_mas}. Around 90 GeV, the signal distribution also shows a bump,
corresponding to the radion production in association with a real Z. We
can eliminate a part of the background by imposing a further cut on the 
missing mass: $m_{mis} > 120$ GeV. 
However, the dominant part of the background remains, and it has  
similar topology with the signal events.   

%----------------------------------------------------------------------
%--------------------------------------------------------------------

\begin{figure}[h]  
\centerline{
\epsfxsize=8.cm\epsfysize=8cm\epsfbox{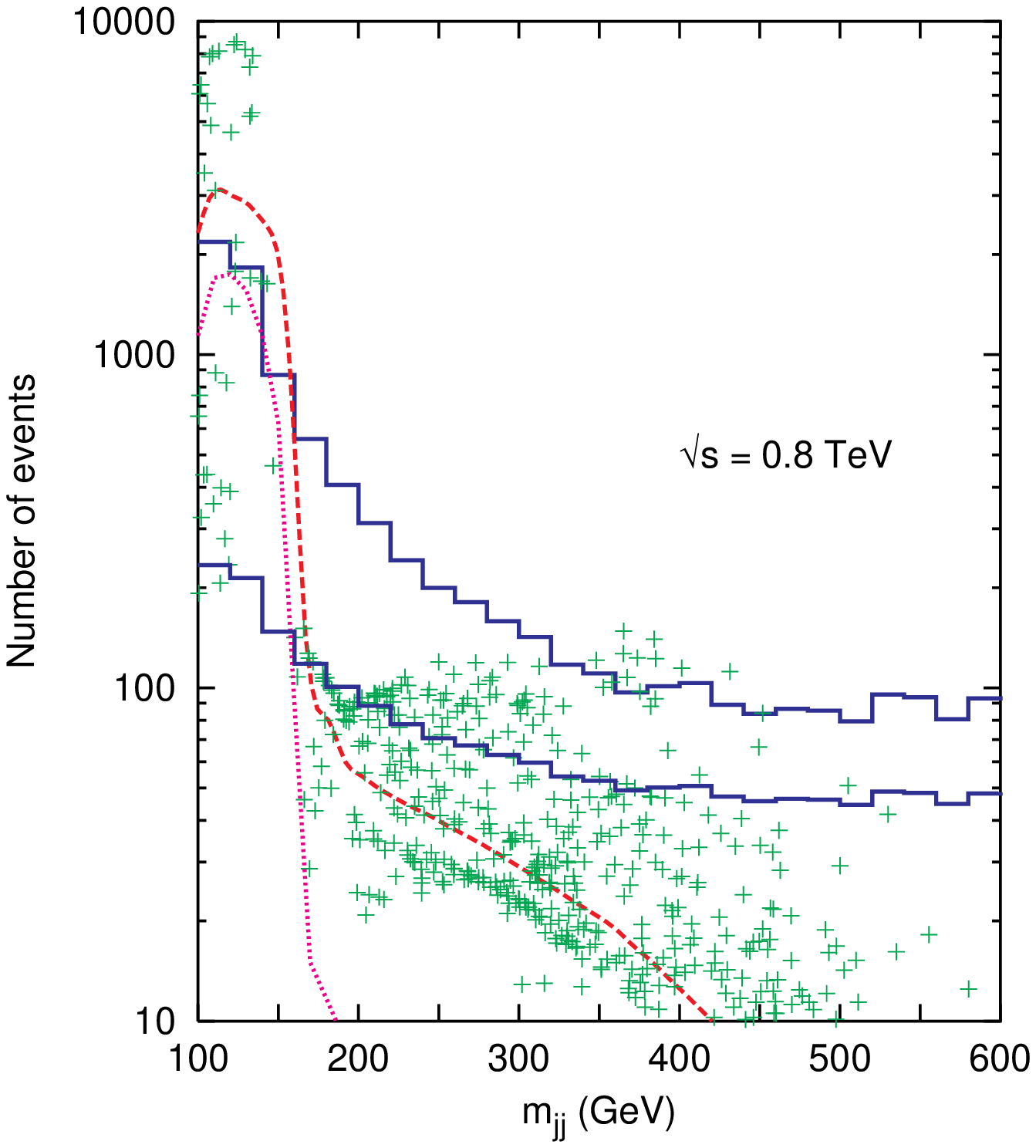}
}
\vspace*{.2cm}
\centerline{
\epsfxsize=8.cm\epsfysize=8cm\epsfbox{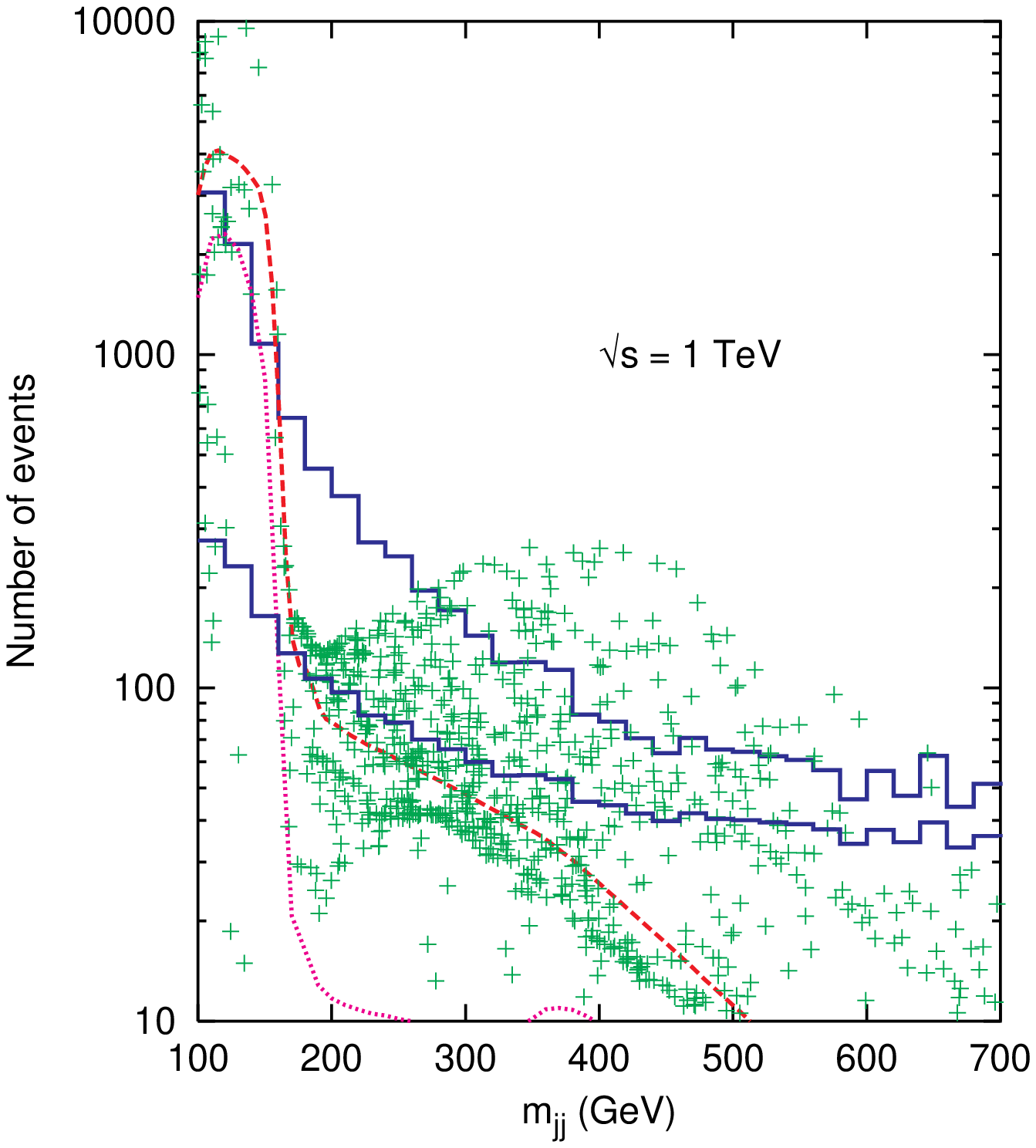}
\epsfxsize=8.cm\epsfysize=8cm\epsfbox{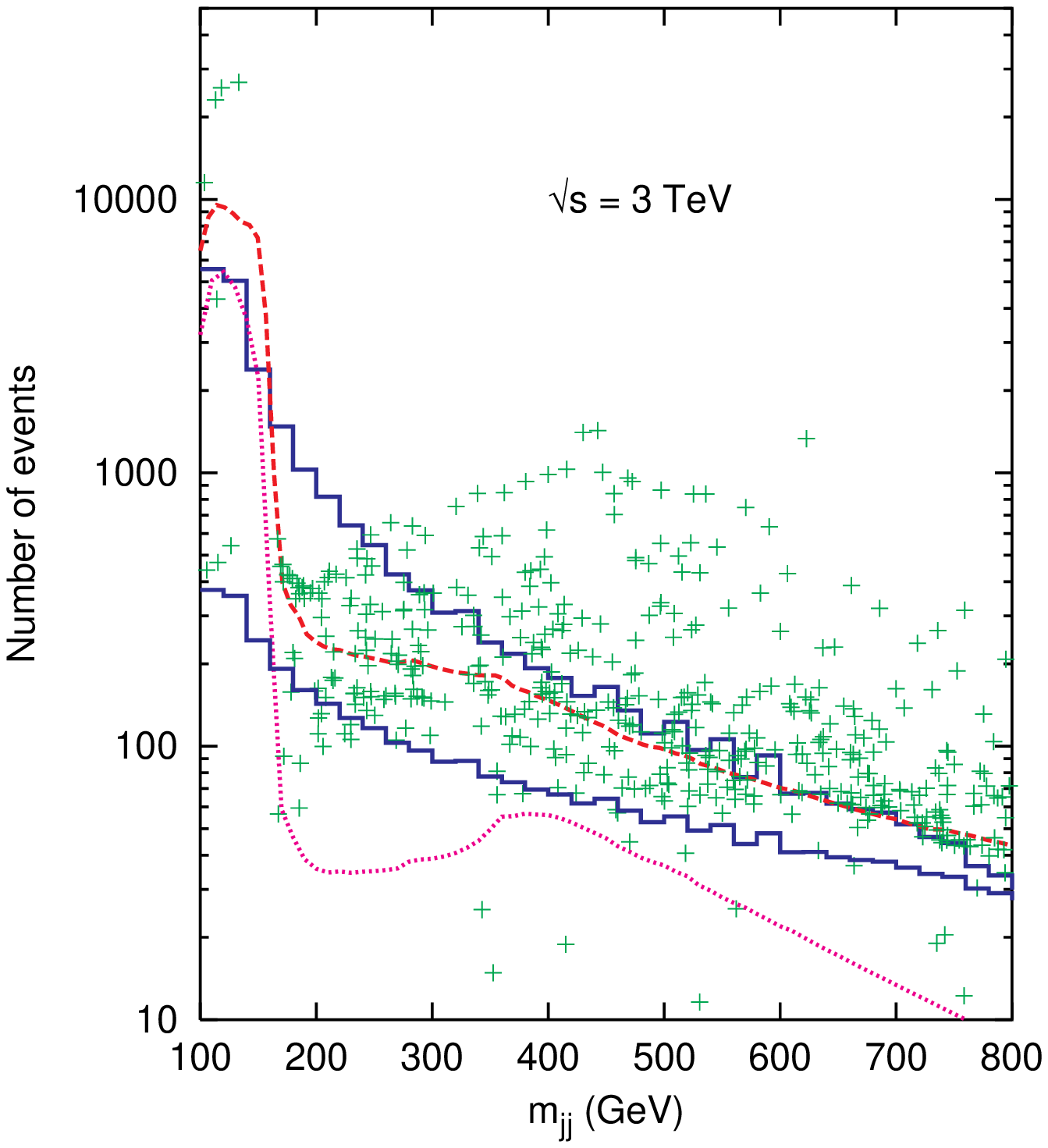}
}
\caption{Number of signal ($+$ for $\xi\ne 0$ and dashed line is for
$\xi=0$), SM higgs (dotted line) and background (solid histogram) events
as a function of the invariant mass of the jets.  The upper histogram
hows the actual number of background events. The lower histogram shows
the 5$\sigma$ fluctuation of background. Points (representing the
signal from higgs and radion) above the lower histogram can be explored 
at 5$\sigma$. Whenever required we fix the higgs mass parameter at 150 GeV.
}
\label{evts}
\end{figure}

In fig. \ref{evts} we have plotted the number of events against the
two-jet invariant mass assuming an integrated luminosity of 500
fb$^{-1}$.  
For radion production  the
two jet invariant mass corresponds to the radion mass.  
When calculating the scatter plots, we have fixed the higgs mass
parameter in the Lagrangian, $m_h=150$ GeV, and we have varied the
radion mass parameter in the range 50 GeV$<m_\phi <$ 800 GeV,
as well as the mixing parameter in the range $-0.75<\xi <0.59$.
For the sake
of comparison we have also plotted the corresponding numbers for the
SM higgs boson (dotted lines), and for the radion in the nonmixed
case, $\xi =0$ (dashed lines).
The upper histogram line shows the
background coming from the processes discussed above.
For the detection the ratio of the signal $S$ to the background $B$ 
should be $S/\sqrt{B} > 5$.
The lower histogram line represents $5\sqrt{B}$.
Thus a $5\sigma$ detection can be done if the parameter space point,
marked by $+$ is above the lower histogram line.
As expected, the cross-sections for both signal and
background grow with the center-of mass energy.  The sudden decrease
of the radion cross-section around $m_{\phi'}=$160 GeV is a reflection of
decrement of $\phi' \rightarrow gg$ branching ratio due to the opening
up of $WW$ decay channel. A local peak in the SM higgs contribution
(only seen in $\sqrt{s} = 3$ TeV panel), around 350 GeV of mass of the
jet pair (same as the higgs mass) is due to the opening up of the 
top-pair threshold to the $h-g-g$ coupling, via top loop.

We can estimate the detection possibilities with radion VEVs other
than 1 TeV, by noticing that the cross-sections behave approximately
as $1/\Lambda_\phi^2$.  This is evident from eqn. 6. The $\phi ' gg$
coupling also depends on $\Lambda_\phi$, but the branching ratio of
$\phi' \rightarrow gg$ is almost insensitive to radion vev. Note that,
apart from an overall $\frac{1}{\Lambda_\phi}$ in the couplings,
quantities like $a_{12},
\theta$ etc. also depends on this vev.  Thus
it is easy to conclude that for $\Lambda_\phi=3$ TeV, a very small
region of the parameter space (mainly for $m_{\phi '} < 200$ GeV)would
still be available with $\sqrt{s}=1$ TeV and a slightly larger region
with $\sqrt{s}=3$ TeV.  $\Lambda_\phi=5$ TeV or above cannot be
detected by any of the studied center of mass energies.

We have not separately presented the numbers for $h'$. However, one can see
from the general arguments that $h'$ production with a neutrino
pair
and the consequent decay to $gg$, is suppressed with respect to the $\phi '$.
%------------------------------------------------------------------
%--------------------------------------------------------------

%------------------------------------------------------------------
\begin{figure}[h]  
\centerline{
\epsfxsize=9.cm\epsfysize=9cm\epsfbox{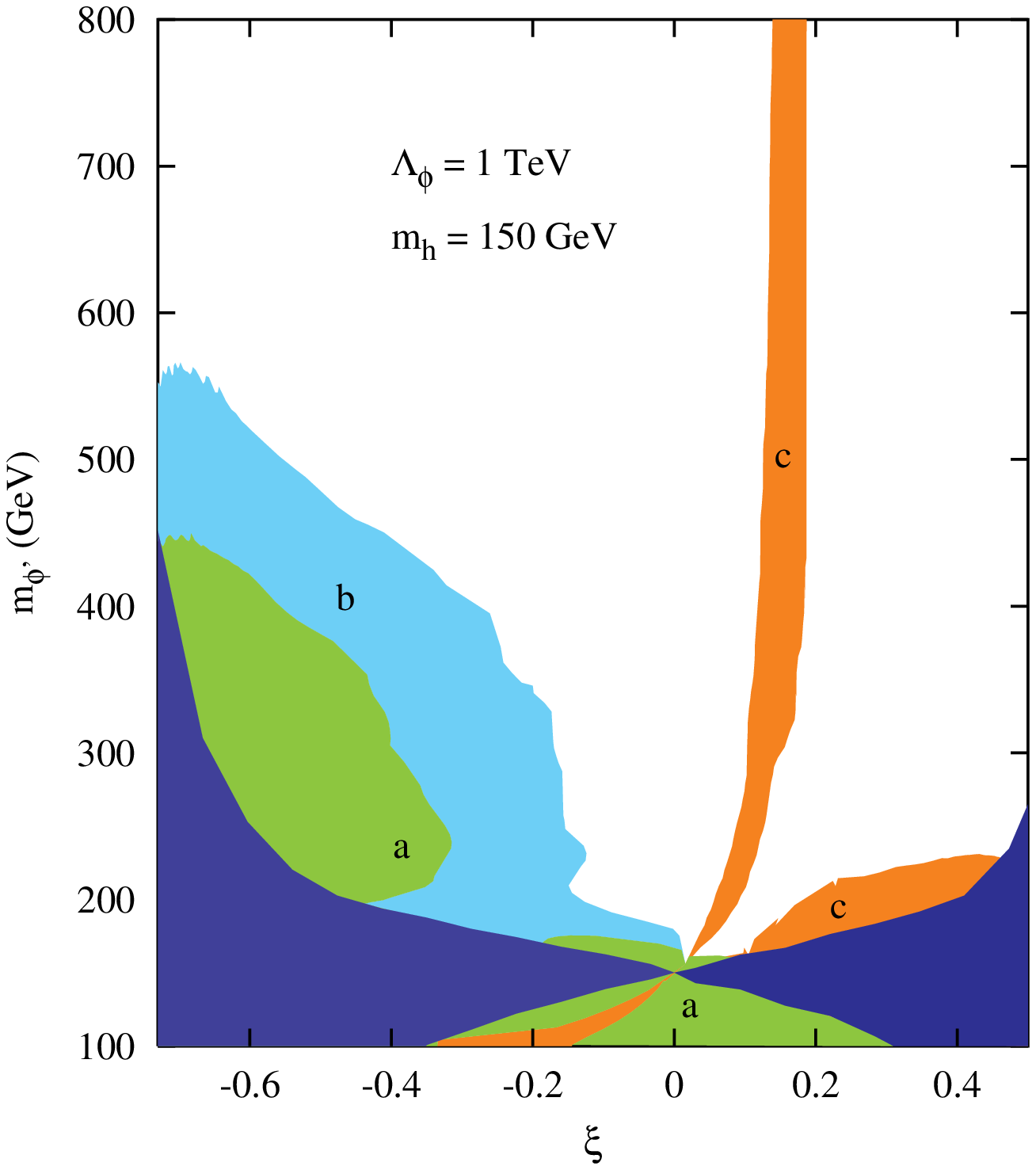}
}

\caption{Radion mass reach as a function of the mixing parameter 
$\xi$. Radion masses, corresponding to the dark wedge-shaped regions
on both sides of $\xi = 0$, are not allowed. Regions (triangular area
close to $\xi = 0$ and in the left side) marked with $a$, can be
probed at $5\sigma$ level with $e^+ e^-$ center of mass energy 800
GeV.  Region marked with $b$ along with $a$ can be probed at $5\sigma$
level with $e^+ e^-$ center of mass energy 1 TeV. The white regions
along with $a$ and $b$ can be probed with 3 TeV $e^+ e^-$ center of mass
energy.  In the shaded regions marked with $c$ (note that there is 
a small
region $c$ close to $\xi = 0$ but for $\xi < 0$) either the radion
production is too small or the $gg$ decay branching ratio is too
small. These regions cannot be probed.
}
\label{contours}
\end{figure}
%--------------------------------------------------------------

It is evident from fig. \ref{couplings} that $h'$ production via $WW$
fusion is less sensitive to $\xi$, than $\phi '$. The production rate is always
suppressed w.r.t the $\phi '$, except for $\xi$ values close to
$\frac{1}{6}$. For $\xi$ values close to 1/6, $\phi '$ branching
ratio to gluons is orders of magnitude larger than the same for
$h'$. We have explicitly checked that number of 2 jet + missing energy
events coming form $h'$ production and decay can only be above the
5$\sigma$ fluctuation of the SM background when higgs mass is below
the $WW$ threshold. Thus, in general one may see two resonances below
the $WW$ threshold, but presence of a single resonance (in gluon gluon
invariant mass distribution) with mass greater than 200 GeV,
definitely points towards the presence of extra scalar like radion.

It is evident from the figures \ref{evts} that radion-higgs mixing
has a very positive effect on the radion search. As for example, with
800 GeV and 1 TeV center-of-mass energy, our proposed signal can test
radion mass up to say 200 GeV in the no-mixing case. When the mixing is
put on, this mass reach is certainly improved for favorable cases. For
all the examples in fig. \ref{evts}, the radion cross-section is well
above the higgs cross-section for $\xi = 0$ case, while non-zero values 
of $\xi$ can push the cross-section in both ways with respect to the 
no-mixing case. 

The above three figures can be nicely summarised in a single plot,
depicting the regions (in $\xi - m_{\phi '}$ plane) in which the
proposed signal is significantly above the background. In
fig. \ref{contours}, we have identified the regions where the numbers
of signal events are more than five times the square root of the
background. The negative values of mixing seem to be favorable to
probe heavier radions for lower center of mass energies. In the
shaded regions marked with $a$ and $b$ along with white region, the
signal strength is higher than the 5$\sigma$ fluctuation of the
background. One can claim a discovery of radions in these regions of
the parameter space. 
Regions marked with $a$ can be probed by a 800 GeV linear collider
and regions marked with $b$ by a 1 TeV linear collider.
To probe the white regions, a 3 TeV linear collider is required.
In the shaded region marked with $c$, (due to
conspiracy of either the radion coupling to $WW/ZZ$ or to $gg$) signal
is too weak compared with the background even at an $e^+ e^-$
center of mass energy of 3 TeV. In fact, in the shaded region (marked
c) parallel to the $m_{\phi '}$ axis, around $\xi = 1/6$, the radion
production is too small. This can be accounted for by the small coupling
of the radion to $WW/ZZ$ as evident from fig. \ref{couplings}. A
larger radion branching ratio to gluons in this region cannot
compensate to yield a signal strength comparable with background.
In the dark wedge shaped regions, no
physical values for radion masses ($m_{\phi '}$) are possible for any
input parameters.  This can be explained from the definition of mixing
angle $\theta$ (see eqn.\ref{mixing}) in terms of other input parameters.

To summarise, we have investigated the possible signature of radions
in Randall-Sundrum scenario at future $e^+e^-$ colliders. If nature
chooses the Randall-Sundrum type of geometry of space and time, then
radion, the lowest lying gravitational excitation in this scenario,
can be explored at the hadronic collider like LHC.  Being a
graviscalar, it couples to the SM particles via the trace of the
energy-momentum tensor. This implies that the couplings are very
similar to that of the SM higgs boson. One important difference from
higgs is the gluon gluon coupling, which is enhanced for radions due
to conformal anomaly.  It is important to look not only for the
possible signatures for this scalar particle at future $e^+e^-$
machines, but also to differentiate this from the SM higgs boson. We
propose to look for the process in which radion is dominantly produced
via $WW$ fusion in $e^+e^-$ collision along with neutrinos and
subsequently decays to a pair of gluons. This in the final state
produces {\em 2 jet {\rm +} missing momentum} signature, with a peak
in two jet mass distribution. We have also estimated the contribution
to this final state from SM higgs boson and also from other SM
backgrounds. Though at an $e^+e^-$ collider with $\sqrt{s}=500$ GeV
the detection is not obvious, at $\sqrt{s}=800$ GeV and above one can
expect $5\sigma$ effect in exploring the radion and differentiating it
from the SM higgs boson.

{\bf Acknowledgments:} Authors thank the Academy
of Finland (project number 48787) for financial support.
 
%%%%%%%%%%%%%%%%%%%%%%%%%%%%%%%%%%%%%%%%%%%%%%%%%%%%%%%%%%%%%%%%%%%%%%%%
%       %%%     Define Journal macros                                  %
% \newcommand{\araa}[3]{{\em Annu. Rev. Astron. Astrophys.\/}
%          {\bf#1} (19#3) #2}                                          %
% \newcommand{\ptp}[3]{{\em Prog. Theoret. Phys. (Kyoto)\/}
% {\if#1} (19#3) #2}                                          %
\newcommand{\plb}[3]{{Phys. Lett.} {\bf B#1} #2 (#3)}                  %
\newcommand{\prl}[3]{Phys. Rev. Lett. {\bf #1} #2 (#3) }        %
\newcommand{\rmp}[3]{Rev. Mod.  Phys. {\bf #1} #2 (#3)}             %
\newcommand{\prep}[3]{Phys. Rep. {\bf #1} #2 (#3)}                   %
\newcommand{\rpp}[3]{Rep. Prog. Phys. {\bf #1} #2 (#3)}             %
\newcommand{\prd}[3]{Phys. Rev. {\bf D#1} #2 (#3)}                    %
\newcommand{\np}[3]{Nucl. Phys. {\bf B#1} #2 (#3)}                     %
\newcommand{\npbps}[3]{Nucl. Phys. B (Proc. Suppl.)
           {\bf #1} #2 (#3)}                                           %
\newcommand{\sci}[3]{Science {\bf #1} #2 (#3)}                 %
\newcommand{\zp}[3]{Z.~Phys. C{\bf#1} #2 (#3)}  
\newcommand{\epj}[3]{Eur. Phys. J. {\bf C#1} #2 (#3)} 
\newcommand{\mpla}[3]{Mod. Phys. Lett. {\bf A#1} #2 (#3)}             %
 \newcommand{\apj}[3]{ Astrophys. J.\/ {\bf #1} #2 (#3)}       %
\newcommand{\jhep}[2]{{Jour. High Energy Phys.\/} {\bf #1} (#2) }%
\newcommand{\astropp}[3]{Astropart. Phys. {\bf #1} #2 (#3)}            %
\newcommand{\ib}[3]{{ ibid.\/} {\bf #1} #2 (#3)}                    %
\newcommand{\nat}[3]{Nature (London) {\bf #1} #2 (#3)}         %
 \newcommand{\app}[3]{{ Acta Phys. Polon.   B\/}{\bf #1} #2 (#3)}%
\newcommand{\nuovocim}[3]{Nuovo Cim. {\bf C#1} #2 (#3)}         %
\newcommand{\yadfiz}[4]{Yad. Fiz. {\bf #1} #2 (#3);             %
Sov. J. Nucl.  Phys. {\bf #1} #3 (#4)]}               %
\newcommand{\jetp}[6]{{Zh. Eksp. Teor. Fiz.\/} {\bf #1} (#3) #2;
           {JETP } {\bf #4} (#6) #5}%
\newcommand{\philt}[3]{Phil. Trans. Roy. Soc. London A {\bf #1} #2
        (#3)}                                                          %
\newcommand{\hepph}[1]{hep--ph/#1}           %
\newcommand{\hepex}[1]{hep--ex/#1}           %
\newcommand{\astro}[1]{(astro--ph/#1)}         %
%       \relax                                                         %
%       %%%     End     Journal macro definitions                      %
%                                            %x
%%%%%%%%%%%%%%%%%%%%%%%%%%%%%%%%%%%%%%%%%%%%%%%%%%%%%%%%%%%%%%%%%%%%%%%%


\begin{thebibliography}{99}
\bibitem{add} N. Arkani-Hamed, S. Dimopoulos and G. Dvali, Phys. Lett.
            {\bf B429 }(1998) 263;
             N. Arkani-Hamed, S. Dimopoulos and G. Dvali, Phys. Lett.
            {\bf B436 }(1998) 257.

\bibitem{rs} L. Randall and R. Sundrum, Phys. Rev. Lett. {\bf 83 }(1999)
             3370;
             L. Randall and R. Sundrum, Phys. Rev. Lett. {\bf 83 }(1999)
             4690.

\bibitem{gold} W.D. Goldberger and M.B. Wise, Phys. Rev. Lett. {\bf 83
}(1999) 4922; W.D. Goldberger and M.B. Wise, Phys. Rev. {\bf D60
}(1999) 107505.

\bibitem{s3a} M. Luty and R. Sundrum, Phys. Rev. {\bf D62} (2000) 035008.

\bibitem{s4} C. Csaki, M. Graesser, L. Randall and J. Terning, 
 Phys. Rev. {\bf D62} (2000) 045015.

\bibitem{s5} W.D. Goldberger and M.B. Wise, Phys. Lett. {\bf B475}
(2000) 275.

\bibitem{radion_pheno}U. Mahanta and S. Rakshit, Phys. Lett. {\bf B480} (2000)
 176; U. Mahanta and A. Datta, Phys. Lett. {\bf B483} (2000) 196;
 K. Cheung, Phys.Rev.{\bf D} 63 (2001) 056007; S. Bae, P. Ko, H.S. Lee
 and J. Lee, Phys. Lett. {\bf B487} (2000) 299; S. Bae, and H.S. Lee,
 hep-ph/0011275; M. Chaichian, A. Datta, K. Huitu, Z.H. Yu,
Phys. Lett. {\bf B524} (2002) 161;
P. Das, U. Mahanta, Nucl. Phys. {\bf B644} (2002) 395; Phys.Lett. 
{\bf B528} (2002) 253, A. Gupta, N. Mahajan, Phys.Rev. {\bf D65} 
(2002) 056003; P. Das, B. Mukhopadhyaya, \hepph{0303135}. 

\bibitem{unitary}T. Han, G. D. Kribs, and B. McElrath,
 Phys. Rev. {\bf D64} (2001) 076003; D.~Choudhury, S.R.~Choudhury, 
A.~Gupta and N.~Mahajan Jour. Phys. {\bf G28}, 1191 (2002).

\bibitem{cheung}K. Cheung, C. S. Kim, J-H. Song, \hepph{0301002}.
\bibitem{gunion} D. Dominici, B. Grzadkowski, J. Gunion and M. Toharia,
\hepph/0206192; Acta. Phys. Polon. {\bf B33} (2002) 2507; M. Battaglia et al.,
\hepph/0304245. 
\bibitem{mixing} G.F. Giudice, R. Rattazzi and J.D. Wells, Nucl. 
Phys. {\bf B 595} (2001) 250.
\bibitem{mixing_1}C. Csaki, M. Graesser and G. Kribs, Phys. Rev. {\bf D63} 
(2001) 065002.
\bibitem{laser} M. Chaichian, K. Huitu, A. Kobakhidze and Z.-H. Yu,
               Phys. Lett.{\bf B} 515 (2001) 65; S.R. Choudhury, A.S. Cornell and
              G.C. Joshi, hep-ph/0012043.

\end{thebibliography}
\end{document}